\documentclass[universe,article,accept,moreauthors,pdftex]{mdpi} 

\firstpage{1} 
\pubvolume{1}
\issuenum{1}
\articlenumber{0}
\pubyear{2022}
\copyrightyear{2022}
\externaleditor{{Academic Editor: Marcello Abbrescia} %MDPI: newly added, please confirm
.
}
\datereceived{29 April 2022} 
\dateaccepted{16 June 2022} 
\datepublished{} 
\hreflink{https://doi.org/} % If needed use \linebreak

\def\msun{M$_\odot$}
\def\iso#1{$^{#1}$}

\newcommand{\apj}{{\it ApJ}}
\newcommand{\apjl}{{\it ApJL}}

\newcommand{\mnras}{{\it MNRAS}}

\newcommand{\prl}{{\it Phys. Rev. Letters}}
\newcommand{\prc}{{\it Phys. Rev. C}}

\newcommand{\gca}{{\it Geochim. Cosmochim. Acta}}

\Title{Origin of Plutonium-244 in the Early Solar System}

\TitleCitation{Origin of Plutonium-244 in the Early Solar System}

\Author{Maria Lugaro %MDPI: Please carefully check the accuracy of names and affiliations.
 $^{1,2,3,4,}$*\orcidA{}, Andr\'es Yag\"ue L\'opez $^{5,\dagger}$\orcidD{}, 
Benj\'amin So\'os $^{1,2,3}$, Benoit C\^ot\'e $^{1,2,6,7,\dagger}$\orcidB{}, M\'aria Pet\H{o} $^{1,2}$\orcidZ{}, \\Nicole Vassh $^{8}$ %MDPI: there is no number 7 for author, please add it before 8.%ML see above
, Benjamin Wehmeyer $^{1,2,9}$ and Marco Pignatari $^{1,2,6,10,\dagger}$}

\AuthorNames{Maria Lugaro, Andr\'es Yag\"ue L\'opez, Benj\'amin So\'os, Benoit C\^ot\'e, M\'aria Pet\H{o}, Nicole Vassh, Benjamin Wehmeyer and Marco Pignatari}

\AuthorCitation{Lugaro, M.; Yag\"ue L\'opez, A.; So\'os, B.; C\^ot\'e, B.; Pet\H{o}, M.; Vassh, N.; Wehmeyer, B.; Pignatari, M.}

\address{%
$^{1}$ \quad Konkoly Observatory, Research Centre for Astronomy and Earth Sciences (CSFK), E\"otv\"os Lor\'and Research Network (ELKH), Konkoly Thege Mikl\'{o}s \'{u}t 15-17, H-1121 Budapest, Hungary; soos.benjamin@csfk.org (B.S.); benoit.cote@csfk.org (B.C.); maria.k.peto@gmail.com (M.P.); benjamin.wehmeyer@csfk.org (B.W.);  	marco.pignatari@csfk.org (M.P.)\\
$^{2}$ \quad CSFK, MTA Centre of Excellence, Konkoly Thege Miklós út 15-17, 1121, Budapest,
 Hungary\\
$^{3}$ \quad Institute of Physics, 
 ELTE E\"{o}tv\"{o}s Lor\'and University, P\'azm\'any P\'eter s\'et\'any 1/A, 1117, Budapest, Hungary 
\\
$^{4}$ \quad School of Physics and Astronomy, Monash University, Clayton, VIC 3800, 
 Australia\\
$^{5}$ \quad Computer, Computational and Statistical Sciences (CCS) Division, Center for Theoretical Astrophysics,\mbox{ Los Alamos National Laboratory,} Los Alamos, NM 87545, USA; ayaguelopez@lanl.gov\\
$^{6}$ \quad Joint Institute for Nuclear Astrophysics---Center for the Evolution of the Elements (JINA-CEE), USA 
\\
$^{7}$ \quad Department %MDPI: the affiliation is not mentioned in author group.
 of Physics and Astronomy, University of Victoria, Victoria, BC V8P 5C2, Canada\\
$^{8}$ \quad TRIUMF, 4004 Wesbrook Mall, Vancouver, BC V6T 2A3, Canada; nvassh@gmail.com\\
$^{9}$ \quad Centre for Astrophysics Research,  
 University of Hertfordshire, College Lane, Hatfield AL10 9AB, UK\\
$^{10}$ \quad Milne Centre for Astrophysics, University of Hull, HU6 7RX, UK
}

\corres{Correspondence: maria.lugaro@csfk.org}

\firstnote{NuGrid Collaboration, \url{http://nugridstars.org}.}

\abstract{We investigate the origin in the early Solar System of the short-lived radionuclide \iso{244}Pu (with a half life of 80 Myr) produced by the $rapid$ ($r$) neutron-capture process. We consider two large sets of $r$-process nucleosynthesis models and analyse if the origin of \iso{244}Pu in the ESS is consistent with that of the other $r$ and $slow$ ($s$) neutron-capture process radioactive nuclei. Uncertainties on the $r$-process models come from both the nuclear physics input and the astrophysical site. The former strongly affects the ratios of isotopes of close mass (\iso{129}I/\iso{127}I, \iso{244}Pu/\iso{238}U, and \iso{247}Pu/\iso{235}U). The \iso{129}I/\iso{247}Cm ratio, instead, which involves isotopes of a very different mass, is much more variable than those listed above and is more affected by the physics of the astrophysical site. We consider possible scenarios for the evolution of the abundances of these radioactive nuclei in the galactic interstellar medium and verify under which scenarios and conditions solutions can be found for the origin of \iso{244}Pu that are consistent with the origin of the other isotopes. Solutions are generally found for all the possible different regimes controlled by the interval ($\delta$) between additions from the source to the parcel of interstellar medium gas that ended up in the Solar System, relative to decay timescales.
If $r$-process ejecta in interstellar medium are mixed within a relatively small area (leading to a long $\delta$), we derive that the last event that explains the \iso{129}I and \iso{247}Cm abundances in the early Solar System can also account for the abundance of \iso{244}Pu. Due to its longer half life, however, \iso{244}Pu may have originated from a few events instead of one only. If $r$-process ejecta in interstellar medium are mixed within a relatively large area (leading to a short $\delta$), we derive that the time elapsed from the formation of the molecular cloud to the formation of the Sun was 9-16 Myr. 
}


\begin{document}
%%%%%%%%%%%%%%%%%%%%%%%%%%%%%%%%%%%%%%%%%%

\section{Introduction}
\label{sec:intro}

There are 17 short-lived %Please check intended meaning has been retained
radioactive (SLR, with~half lives of the order of 0.1 to 100 Myr) nuclei known to have been (or potentially have been) present in the early Solar System \mbox{(ESS) ~\cite{lugaro18rev}. }Three of them, \iso{129}I, \iso{244}Pu, and~\iso{247}Cm, have the specific property to be produced in the Galaxy almost exclusively by the process of $rapid$ neutron captures (the $r$ process). Among~those three, live \iso{244}Pu from the present-time interstellar medium has also been detected in young sediments of the ocean floor~\cite{wallner15,wallner21}. Furthermore, \iso{244}Pu and \iso{247}Cm are actinides located beyond Pb and Bi at mass numbers around 208-210, the~end point of the $slow$ neutron-capture ($s$) process~\cite{ratzel04}. Therefore, they are exclusively of $r$-process origin. Being located beyond the classical third $r$-process peak at Pt and Au, actinides are typically produced if the number of neutrons per seed is relatively large. 
%such as neutron star - neutron star (NS-NS) or neutron star - black hole (NS-BH) mergers  \citep{Rosswog2013}.
Instead, \iso{129}I belongs to the classical second $r$-process peak. It has only a very minor (a few percent) contribution from the $s$ process because the unstable isotope that precedes it on the $s$-process path, \iso{128}I, has a half life of 25 min only and decays faster than the typical time required for capturing a neutron. 
%Therefore the bulk of \iso{129}I in the Galaxy constitutes the lower mass peak of the r-process.  The proposed production sites of \iso{129}I are a) magnetorotationally driven supernovae \citep{Winteler2012} and/or b) compact binary mergers.  
These three $r$-process isotopes have most likely the same $r$-process origin (as indicated by elemental abundance observed in halo stars,~\cite{cowan21}). They can be studied individually or together to provide evidence on the history of the material that made up the Solar System~\cite{lugaro14science} and to set constraints on the $r$-process astrophysical site and its nuclear input, which are both extremely uncertain~\cite{hotokezaka15,cote21science,wang21a,wang21b}. In~particular, C\^ot\'e~et~al.~\cite{cote21science} (hereafter Paper I) constrained the last $r$-process source to have contributed to the solar material by comparing the \iso{129}I/\iso{247}Cm ratio observed in primitive meteorites to nucleosynthesis calculations based on neutron star (NS-NS) merger, black hole--neutron star (NS-BH) merger, and~magneto-rotational supernova simulations. Here, we extend that study to \iso{244}Pu, to~investigate if it is possible to find an explanation for the presence of this SLR isotope in the ESS compatible with the explanation for the presence of the other SLR isotopes heavier than iron, also well known to have been present in the ESS.
%and to better constrain the timing of the last event that contributed r-process nuclides to the material building up our Solar System. 

Table~\ref{tab:intro} summarises the main properties and information available on the four isotopic ratios under consideration here: \iso{129}I/\iso{127}I, \iso{244}Pu/\iso{238}U, \iso{247}Cm/\iso{235}U, and~\iso{129}I/\iso{247}Cm. We only analyse isotopic ratios because the most direct evidence that comes from the analysis of meteoritic material on ESS values is not absolute abundances, but~abundance values relative to each other. Absolute abundances suffer from many uncertainties, e.g.,~chemical separation in the nebula, in~the meteorite parent body, and/or during chemical analysis, as~well as dilution from the original stellar source. The~ratios of interest are those of each estimated SLR abundance relative to a long-lived or a stable isotope. These ratios are directly measured in primitive meteorites and their components (the first three rows of Table~\ref{tab:intro}), or derived from the ratios directly measured (last row), as~in the case of the \iso{129}I/\iso{247}Cm ratio. This last ratio provides us with a further observational constraint because \iso{129}I and \iso{247}Cm have very similar half lives~\cite{yague21PaperIII}\endnote{The mean-life $\tau_{ratio}$ of the \iso{129}I/\iso{247}Cm ratio given in the table was obtained by Monte Carlo sampling of the uncertainties on the mean lives of the two isotopes, $\tau_{129}$ and $\tau_{247}$, which are 5\% and 6\%, respectively, at~2$\sigma$ (for comparison, the uncertainty for \iso{244}Pu is 2\%) within the usual formula: $\tau_{129} \times \tau_{247} /( \tau_{129} -\tau_{247})$. Using the recommended values, $\tau_{ratio}$ would be equal to 2449 Myr, however, sampling of the uncertainties produces a lower value most of the time because the uncertainties make $\tau_{129}$ and $\tau_{247}$ move away from each other, and~therefore their difference, at~denominator in the formula above, increases. 
In general, it would be extremely useful if the half lives of \iso{129}I and \iso{247}Cm  could be measured with higher precision than currently available. 
A more detailed statistical analysis should also be carried out considering that 
the peak value reported in the table is probably not the best statistical choice due to the exponential behaviour of the decay. In~fact, although~$\tau \sim 270$ Myr is the most common value, when $\tau \gtrsim 1000$ Myr abundances do not vary much anymore within the time scales, roughly $< 200$ Myr are %Please check intended meaning has been retained
of interest for the ESS (discussed in Section~\ref{sec:galaxy}). Therefore, a~more statistically significant value may be higher than the peak value reported in the table, probably around 900 Myr.
For the other ratios, the~values of the mean lives at numerator and denominator in the equation above are so different that $\tau_{\rm ratio}$ is always within 2\% of the $\tau$ of the short-lived isotope. A~statistical analysis of the uncertainties would not affect those values, although~we will analyse statistically the impact of the uncertainties on all the mean lives when we derive timescales in Section~\ref{sec:galaxy}.}.
This allowed to remove several theoretical uncertainties in Paper I, providing a direct window into the astrophysical conditions of the $r$-process site that produced the \iso{129}I and \iso{247}Cm in the ESS.
Note that, instead, it is not possible to extract any further meaningful constraints from the  \iso{129}I/\iso{244}Pu and \iso{247}Cm/\iso{244}Pu ratios because their half lives are very different from each other~\cite{yague21PaperIII}.

\begin{table}[H]
\caption{Properties 
%MDPI: Please make sure that permission has been obtained and there is no copyright issue.%ML OK, and changed to Properties as it is a better word here
 of the three ratios that involve SLR nuclei of $r$-process origin that were present in the ESS: the mean lives of the isotopes at numerator, at~denominator, and~of their ratio ($\tau_{\rm num}$, $\tau_{\rm den}$, and~$\tau_{\rm ratio} = 
\tau_{\rm num} \times \tau_{\rm den} /( \tau_{\rm num} -\tau_{\rm den})$, respectively, all in Myr), and~the ESS values (at 2$\sigma$, from~\cite{lugaro18rev}). We also show, in the last column, the three values of the $K$ factor that affect each of the ratios when predicted by the GCE model. This factor accounts for the star formation history and efficiency, the~star-to-gas mass ratio, and~the galactic outflows (Section~\ref{sec:galaxy}). The~uncertainties on these quantities result in a minimum ($K_{\rm min}$), a~best-fit ($K_{\rm best}$), and~a maximum ($K_{\rm max}$) value of each ratio.
\label{tab:intro}}
\newcolumntype{C}{>{\centering\arraybackslash}X}

\begin{adjustwidth}{-\extralength}{0cm}
%\centering %% If there is a figure in wide page, please release command \centering
\begin{tabularx}{\fulllength}{CCCCCC}
\toprule
%\begin{tabular}{cccccc}
%\textbf{SLR nucleus} & \textbf{$T_{1/2}$} & \textbf{$\tau$} & \textbf{reference} & \textbf{ESS ratio} & \textbf{$K_{\rm min}$, $K_{\rm best}$, $K_{\rm max}$} \\
\textbf{Ratio} & \boldmath{\textbf{$\tau_{\rm num}$}} &  \boldmath{\textbf{$\tau_{\rm den}$}} & \boldmath{ \textbf{$\tau_{\rm ratio}$}} & \textbf{ESS Ratio} &  \boldmath{\textbf{$K_{\rm min}$, $K_{\rm best}$, $K_{\rm max}$}} \\
\midrule
%\iso{129}I & 15.7(0.8) & 22.6 & \iso{127}I & ($1.28\pm0.03)\times 10^{-4}$ & 1.6, 2.3, 5.7 \\
\iso{129}I/\iso{127}I & 22.6 & stable & 22.6 & ($1.28\pm0.03)\times 10^{-4}$ & 1.6, 2.3, 5.7 \\
\iso{244}Pu/\iso{238}U & 115 & 6447 & 117 & ($7\pm2)\times 10^{-3}$ & 1.5, 1.9, 4.1 \\
\iso{247}Cm/\iso{235}U & 22.5 & 1016 & 23.0 & ($5.6\pm0.3)\times 10^{-5}$ & 1.1, 1.2, 1.8 $^b$ \\
% 1.4, 1.9, 3.2 in the science paper SM ... 
%"Yes, 1.5 years ago (~25/9/2020) we found out that the numbers for K_U were wrong." see Andres's email 17 Feb 2022
%\hline
\midrule
\iso{129}I/\iso{247}Cm & 22.6 & 22.5 & 270 $^a$ (100--3000) &  $438\pm184$ & 1, 1, 1 \\
\bottomrule
\end{tabularx}
\end{adjustwidth}
$^a$ Values taken from the asymmetric $\tau_{\rm ratio}$ distribution shown in Figure~S4 of Paper I. The~first value is roughly the peak of the distribution, and~the values in parenthesis represent most of its total range. $^b$ Values corrected relative to those reported in Paper I.
\end{table}
%\unskip

Out of the four ratios reported in Table~\ref{tab:intro}, \iso{244}Pu/\iso{238}U has not been considered yet within a global analysis of origin of the SLR nuclei heavier than iron in the ESS. This is for two main reasons: first, its half life of 80 Myr is very different from that of the other two isotopes of roughly 15 Myr, therefore, the~modelling of its abundance in the interstellar medium (ISM) is likely to present a different behaviour (see discussion in Section~\ref{sec:galaxy}). Second, its ESS abundance is less certain than those of the other two isotopes. The~ESS \iso{129}I/\iso{127}I ratio has an uncertainty of roughly 2\% at 2$\sigma$ and many studies agree on its value, suggesting that systematic uncertainties are not significant~\cite{gilmour06}. The~\iso{247}Cm/\iso{235}U was established with an uncertainty of roughly 6\% at 2$\sigma$ thanks to the discovery of the special meteoritic inclusion, named Curious Marie, rich in U~\cite{tissot16}. More data on different samples is still needed to completely establish this~value. 

In the case of the ESS \iso{244}Pu abundance (i.e., the~\iso{244}Pu/\iso{238}U ratio), instead, not only is the uncertainty for the value reported in Table~\ref{tab:intro} roughly 30\%, but~also there are potential systematic uncertainties in the determination of the ESS value. The~ESS \iso{244}Pu abundance can be estimated by xenon isotope studies of meteorites, since \iso{129}Xe and the heavy \iso{131-136}Xe are stable isotopes produced by the spontaneous fission of \iso{244}Pu. Moreover, solids are extremely poor in noble gases and the radiogenic and fissiogenic xenon signatures may become significant over time and, hence, can be quantified at high precision. Studies have been focusing on gas-poor meteoritic materials: mineral separates \citep{Wasserburg1969PhRvL}; CAIs \citep{Marti1977LPI, Podosek1972E&PSL};  differentiated meteorites with simple cooling histories, such as angrites  \citep{Lugmair1977E&PSL}) and\mbox{ eucrites  \citep{Shukolyokov1996GeCoA};} and high-metamorphic-grade ordinary chondrites  \citep{hudson89}. Currently, there are two ``best estimates'' of ESS using different approaches. \citet{Lugmair1977E&PSL} normalized \iso{244}Pu to \iso{150}Nd, an~$r$-process-only isotope of Nd, because~they found an achondrite (Angra dos Reis) where they could prove that the geochemical analogue of Pu is Nd, and~potential modification of the Pu/Nd ratio with respect to the Solar System abundances can be ruled out. They reported \iso{244}Pu/\iso{238}U ratios $\simeq$0.0043 at the adjusted time of Solar-System formation \citep{connelly12}.  The~value reported in Table~\ref{tab:intro} (0.007) is a different estimate by~\cite{hudson89}, who used a different approach. As~the fissiogenic signature is dominated by \iso{244}Pu-derived xenon in meteorites, they irradiated an exceptionally gas-poor ordinary chondrite (St Severin) with thermal neutrons to induce the fission of \iso{235}U and derived the \iso{244}Pu/\iso{238}U ratio from the component analysis of xenon isotope measurements alone. This value is almost twice as high as the value provided by the Angra dos Reis study, and~it is in better agreement with the more recent analysis of Xe in ancient terrestrial zircons from Western Australia~\cite{turner07}. In~summary, the~major challenge is to find a meteorite sample that is representative of the Solar System, and~for which geochemical processes that could potentially modify the relative abundances of Pu to U or rare earth elements with respect to the chondritic composition is well-understood, and~the effect can be corrected for.  
Here, we will consider for the ESS \iso{244}Pu/\iso{238}U ratio the value reported in Table~\ref{tab:intro}. If~the ``true'' value was eventually found to be lower, for~example, by~a factor of two, all the times calculated and reported in our analysis below would have to be increased by 80~Myr.

%Overall, the main difficulty in determining the ESS \iso{244}Pu/\iso{238}U ratio is thepresence of variations in the chemical behaviour of U and Pu as material is condensed into the solid phase and then processed into the minerals where the decay products can be measured.

The aim of this paper is to investigate possible self-consistent solutions for the origin of the abundances of all the SLR nuclei heavier than iron observed to have been present in the ESS, including \iso{244}Pu. These observed abundances are represented by the four $r$-process ratios reported in Table~\ref{tab:intro}, as~well as the SLR isotopes produced by $slow$ neutron captures (the $s$ process, specifically \iso{107}Pd, \iso{135}Cs, and~\iso{182}Hf, as~discussed in~\cite{trueman22}). We start by discussing predictions from state-of-the-art models of the $r$ process for the three SLR nuclei of interest and their reference isotopes (Section~\ref{sec:yields}). Then, in Section~\ref{sec:galaxy}, we consider the temporal evolution of the \iso{244}Pu/\iso{238}U ratio in the ISM and discuss if there are solutions for its ESS value that are consistent with the abundances of the other SLR nuclei heavier than iron. Finally, in Section~\ref{sec:conclusions}, we present our summary, conclusions, and~suggestions for future~work. 
 
%In spite of this difficulty, we believe that investigation of \iso{244}Pu abundance in the ESS and presentation of its production and its yields is timely also given the recent detection of this nucleus of interstellar origin live in Earth's crust samples~\cite{wallner15,wallner21}.

\section{Nucleosynthesis~Calculations}
\label{sec:yields}

We consider the large set of $r$-process abundances published with Paper I and calculated with the nuclear network code WINNET%MDPI: please check the link, if~here should add accessed date.%ML added
\endnote{\url{https://zenodo.org/record/4446099\#.YgKVxWAo-mk}  (accessed on 15 June 2022)} ~\cite{marius_yields,Winteler2012} and the nucleosynthesis network \hl{PRISM}%MDPI: please check the link, if~here should add accessed date..
\endnote{\url{https://zenodo.org/record/4456126\#.YgKV0GAo-mk}  (accessed on 15 June 2022)} ~\cite{nicole_yields,mumpower18}. 
All the abundances reported and~used in this work, are taken at 1 Myr after the nucleosynthetic event, i.e.,~they are not decayed completely, given that we are interested in SLR nuclei. 
%The choice of 1 Myr is required to make sure that all the shorter-lived isotopes that contribute to the abundance of the isotopes considered here are decayed. At the same time, we keep the effect of time minimal on the abundances of the 6 isotopes considered here, with a 5\% change at most in original abundance of \iso{129}I and \iso{247}Cm.

%\unskip
%Total mass ejected from Marius's models:
%R1010: 7.64e-3 Msun
%Bs125: 5.5e-4 Msun
%Wmhd: 6.72e-3 Msun
%FMdef: 1.701e-3 Msun

Table~\ref{tab:WINNET} lists all the WINNET models considered here and the relationship between the labels used in Paper I and the shorter labels used here. The~sites and the nuclear physics sets, with~all their relevant references, are described in detail in Paper I and Ref.~\cite{eichler19}. Here, we remind briefly that the nomenclature of the nuclear input is as follows: [D,J,Jm] denotes the mass model (D for Duflo Zuker, J for JINA reaclib, Jm for JINA with Marketin theoretical $\beta$ decays). The~``h'' indicates that the nuclear heating subroutine was turned on~\cite{freiburghaus99}), modifying the temperature evolution of the trajectory. Finally, [f1,f2,f4] represent three different fission fragment models.
There are, in total, 3 (top)$\times$3 (bottom) nuclear labels, i.e.,~nine sets of nuclear inputs (right side of the table), and~seven astrophysical sites (left side of the table), therefore, a~total of 63 WINNET models. The~tabulated abundances of the six isotopes of interest here (together with the Eu isotopes and \iso{232}{Th} for future reference) can be found in Supplementary Table~S1.

\begin{table}[H]
\caption{The correspondence %MDPI: please check the vertical line, if~it is necessary? %ML: yes
 of the astrophysical site and nuclear input labels are used here to indicate the WINNET models and %Please check intended meaning has been retained
 those used in Paper I, where a full description of each site and nuclear model and relevant references can also be found. The~total mass ejected by each site is also indicated. \label{tab:WINNET}}
%\newcolumntype{C}{>{\centering\arraybackslash}X}
\begin{tabularx}{\textwidth}{CCC|CC}
\hline
%\begin{tabular}{cccccc}
%\textbf{SLR nucleus} & \textbf{$T_{1/2}$} & \textbf{$\tau$} & \textbf{reference} & \textbf{ESS ratio} & \textbf{$K_{\rm min}$, $K_{\rm best}$, $K_{\rm max}$} \\
\textbf{Site Label} & \textbf{Site Label (Paper I)} & \textbf{Mass Ejected (\msun)} & \textbf{Nuclear Label} & \textbf{Nuclear Label (Paper I)} \\
\hline
R1010 & NS-NS merger dyn. ejecta (R) & $7.64\times 10^{-3}$ & Dhf & DZ10 \\
R1450 & NS-BH merger dyn. ejecta (R) & $2.38\times 10^{-2}$ & Jhf & FRDM \\
Bs125 & NS-NS merger dyn. ejecta (B) & $5.50\times 10^{-4}$ & Jmhf & FRDM(D3C*) \\
FMdef & NS-NS merger disk ejecta 1 & $1.70\times 10^{-3}$ & 1 & Panov \\
FMs6 & NS-NS merger disk ejecta 2 & $1.27\times 10^{-3}$ & 2 & K \& T \\
FMv0.10 & NS-NS merger disk ejecta 3 & $4.06\times 10^{-3}$ & 4 & ABLA07 \\
Wmhd & MR SN & $6.72\times 10^{-3}$ & & \\
\hline
\end{tabularx}
\end{table}

In the case of the PRISM models, the~nomenclature is identical to that used in Paper I (Table S3), and~references therein. In~this case, four sites are considered (the dynamical ejecta of two NS--NS mergers and two NS--BH models), and~combined with ten different mass models, of~which four are also investigated using alternative $\beta$ decays (the ``Mkt'' label, here, corresponding to the ``D3C*'' label in Paper I). The~total is, therefore, 4 $\times$ 14 = 56 PRISM models. The~tabulated ratios of interest here can be found in Supplementary Table~S2.

\begin{figure}[H]
\begin{adjustwidth}{-\extralength}{0cm}
\centering %% If there is a figure in wide page, please release command \centering
\includegraphics[width=14 cm]{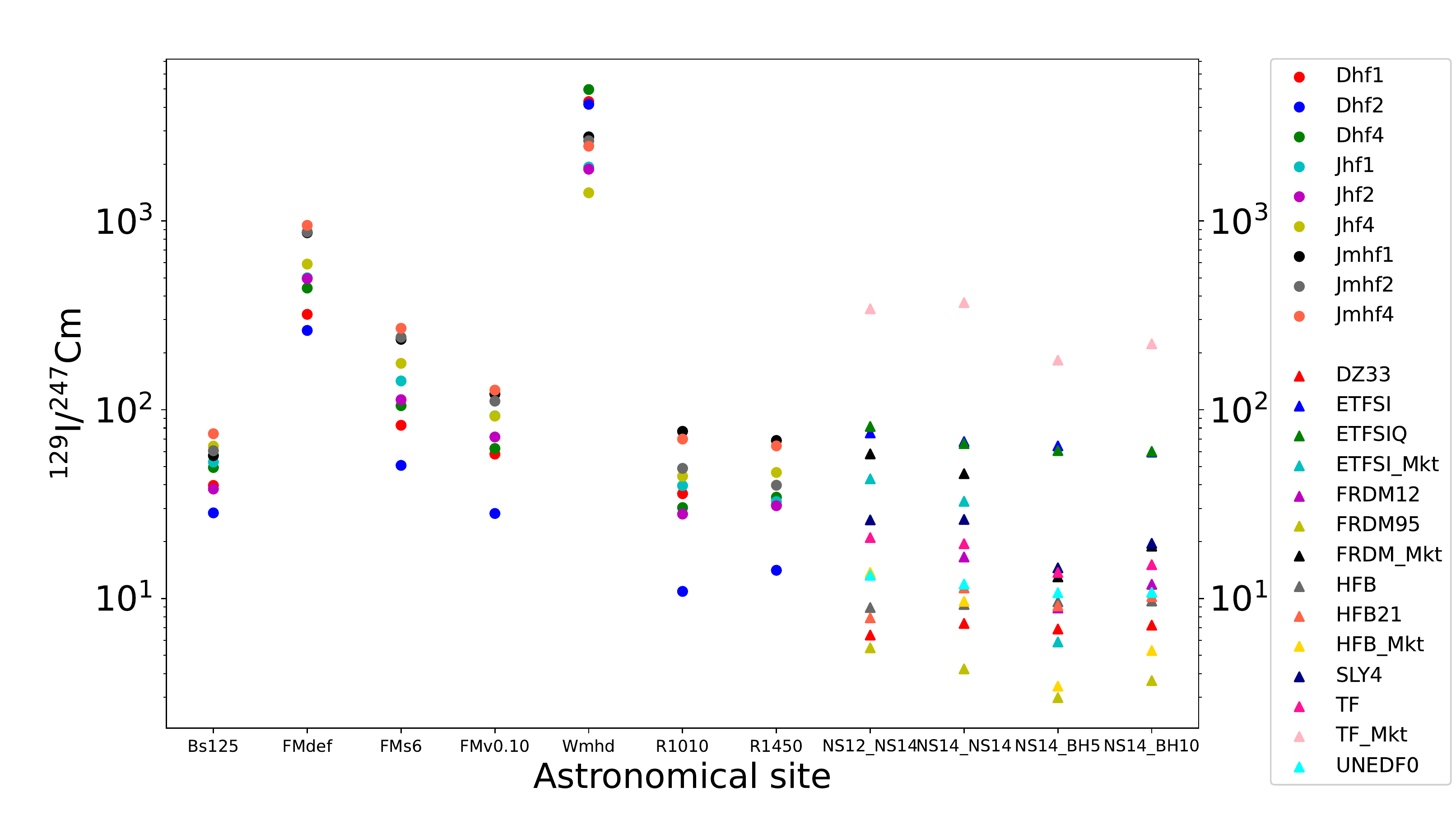}
\end{adjustwidth}
%\begin{figure}[H]
%\includegraphics[width=9.5 cm]{lastevent/FMs6 Fmhf1_best.pdf}
\caption{Ratios of the \iso{129}I/\iso{247}Cm isotopic ratio from all the WINNET (circles) and PRISM (triangles) models, with~color as indicated in the box on the right side. The~detailed label description can be found in Table~\ref{tab:WINNET} and in Paper I. Note that the scale is~logarithmic. \label{fig:ratios129247}}
\end{figure}

\begin{figure}[H]
\begin{adjustwidth}{-\extralength}{0cm}
\centering %% If there is a figure in wide page, please release command \centering
\includegraphics[width=12 cm]{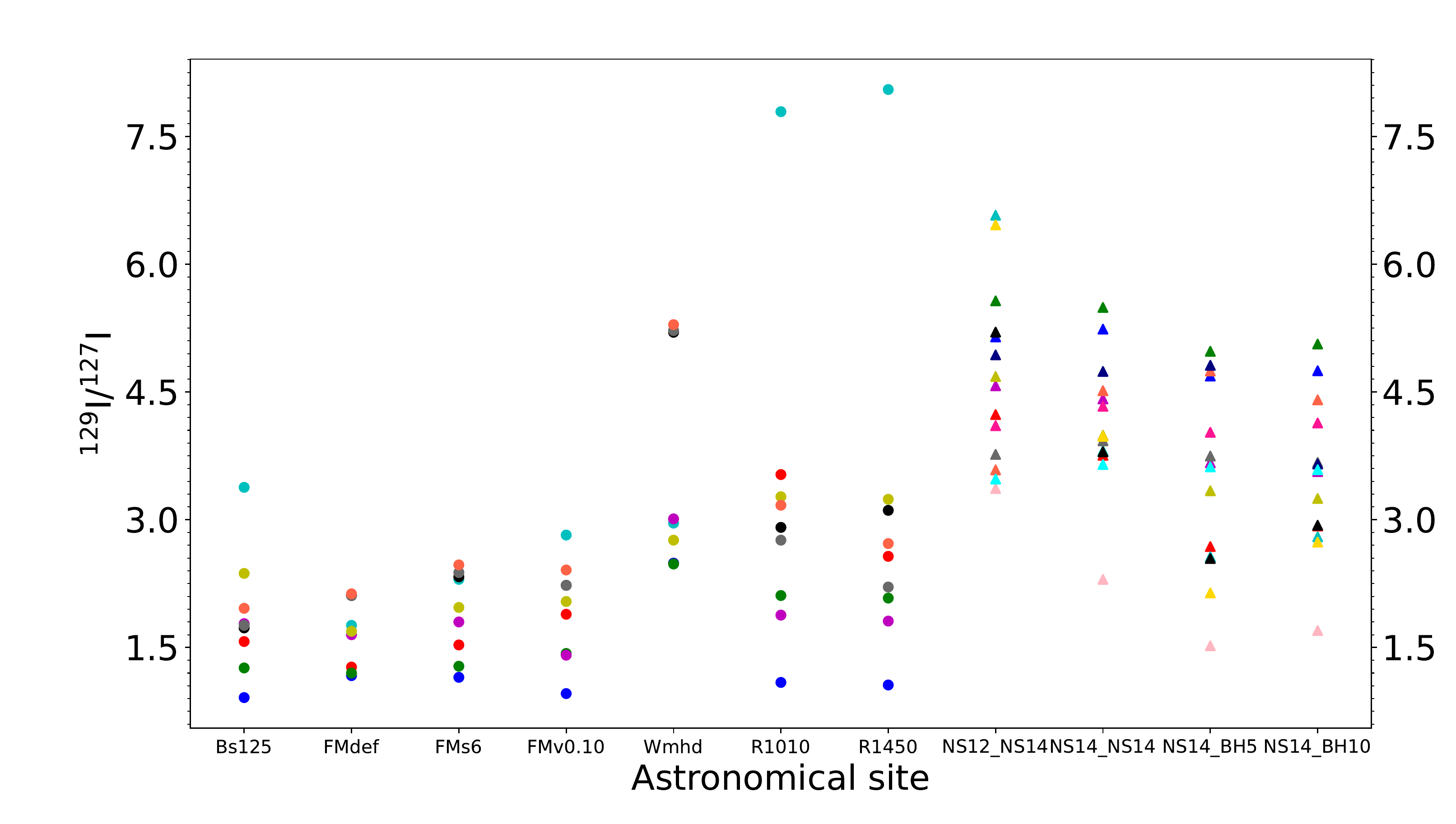}
\includegraphics[width=12 cm]{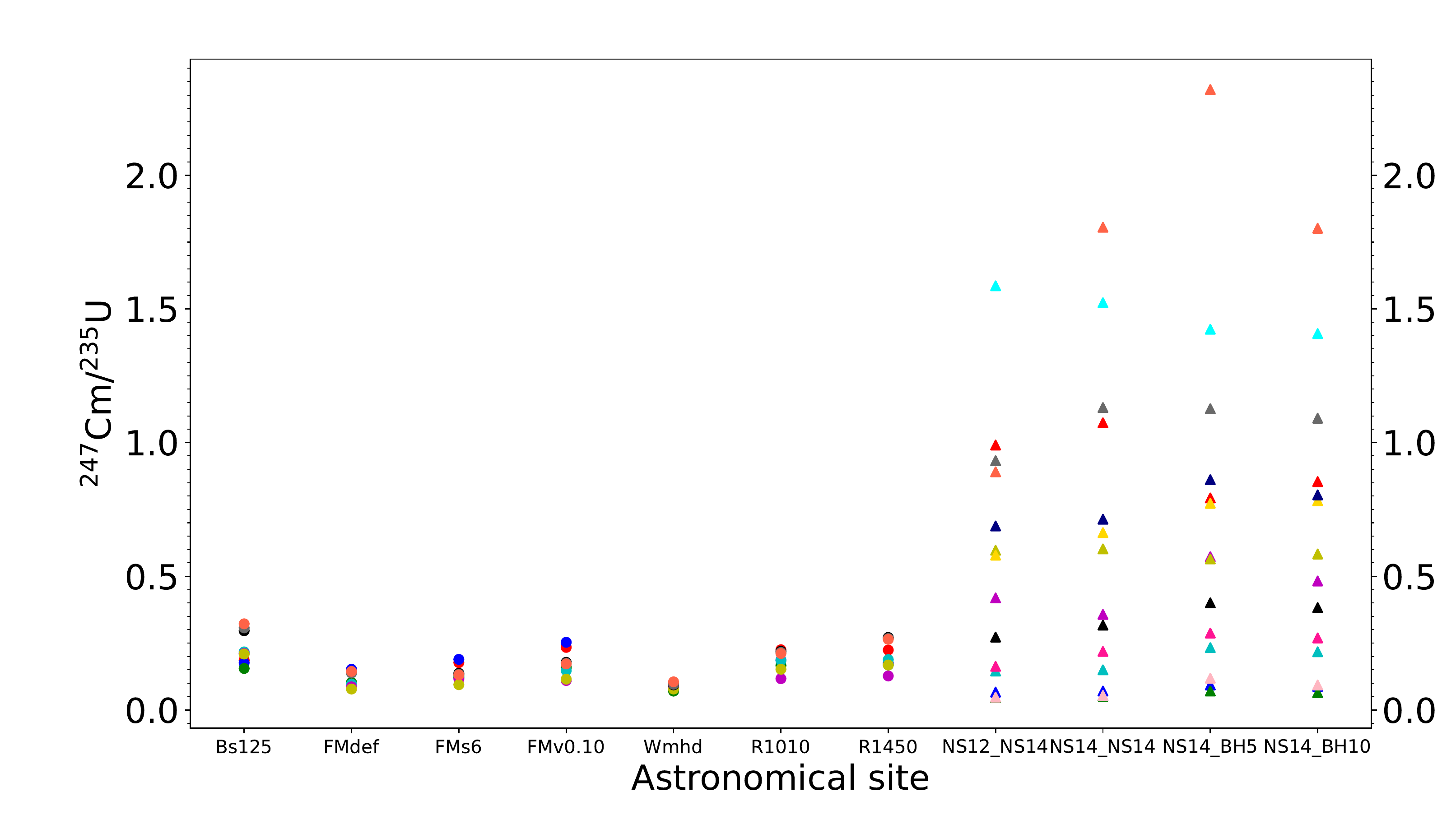}
\includegraphics[width=12 cm]{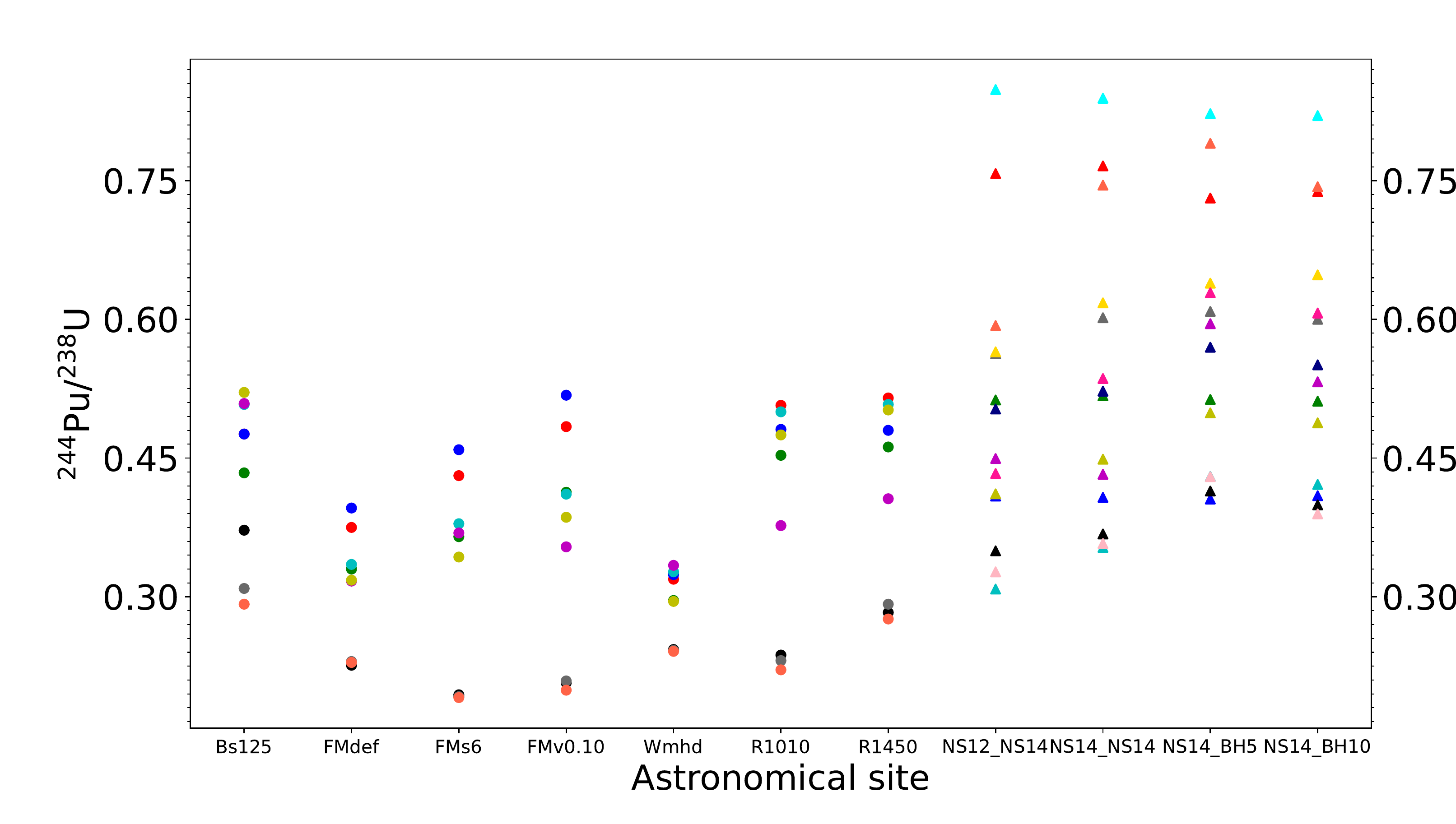}
\end{adjustwidth}
%\begin{figure}[H]
%\includegraphics[width=9.5 cm]{lastevent/FMs6 Fmhf1_best.pdf}
\caption{Ratios of the abundances of the three SLR of interest to their stable or long-lived isotopes from all the WINNET (circles) and PRISM (triangles) models, with~colors as in Figure~\ref{fig:ratios129247}. Note that, differently to Figure~\ref{fig:ratios129247}, all the scales here are~linear. \label{fig:ratios}}
\end{figure}

The four ratios of interest from all the models are plotted in Figure~\ref{fig:ratios129247} and \ref{fig:ratios}. As~expected, ratios of isotopes of similar mass (Figure~\ref{fig:ratios}) are much less dependent on the model than the \iso{129}I/\iso{247}Cm ratio (Figure~\ref{fig:ratios129247}). Variations in those three ratios are typically of factors $\sim$2.5 to 3 in the WINNET models. Additionally found by some of these models is the \iso{129}I/\iso{127}I ratio of 1.35 derived from the $r$-process abundance of the stable \iso{129}Xe in the Solar System\endnote{This is calculated using the residual method, where the $r$-process abundance is the total solar abundance of \iso{129}Xe minus the predicted $s$-process abundance. This method cannot be applied to \iso{247}Cm and \iso{244}Pu as these isotopes do not have one daughter stable nucleus produced exclusively by their decay.}. 
When considering the PRISM models, which explored a larger set of nuclear inputs, variations are somewhat larger, especially in the case of \iso{247}Cm/\iso{235}U (up to a factor of 10). 

All the models show variations in the \iso{129}I/\iso{247}Cm ratio of up to three orders of magnitudes (Figure~\ref{fig:ratios129247}, corresponding to Figure~S2 of Paper I, but~with the PRISM models also included). Of~the WINNET models, 13 of them (the nine FMdef models, the~three FMs6 Jmhf models, and~the FMs6 Jhf4 model) could match the observed \iso{129}I/\iso{247}Cm in the ESS. The~PRISM models all represent dynamical ejecta and, therefore, provide similar results to the corresponding WINNET models (Bs125, R1010, R1450). As~noted above, the PRISM models explore a larger set of nuclear inputs and only one out of those 14 choices (TF\_Mkt) provides a solution for the \iso{129}I/\iso{247}Cm ratio in all the four~sites.

A %MDPI: we remove the bold format, please check. %ML: thank you!
 quick estimate indicates that self-consistent values of the time elapsed from the last 
$r$-process event (Section~\ref{sec:last}) would result from models with similar values of the 
\iso{129}I/\iso{127}I and \iso{247}Cm/\iso{235}U ratios. This is because both the ESS value and 
the $K$ values for \iso{129}I/\iso{127}I are roughly twice those of \iso{247}Cm/\iso{235}U. 
Therefore, the~two differences cancel each other out in the calculation of the ISM ratio needed 
to derive the time interval of the decay by comparison to the ESS ratio\endnote{When we also 
consider the difference due to fact that while \iso{127}I is stable, \iso{235}U will also 
decay. For~time intervals of the order of 100-200 Myr, this corresponds to a small effect on 
the \iso{247}Cm/\iso{235}U ratio of roughly 10-20\%.}. While there are no models with the same 
values of the \iso{129}I/\iso{127}I and \iso{247}Cm/\iso{235}U ratio, when we consider the 
uncertainties of the $\tau$ and ESS values, many solutions can be found for a much larger range 
of relative ratios, as~shown in Section~\ref{sec:last}. This is because the time interval of 
the decay is a function of the natural logarithm of the abundance ratio; therefore, variations 
in the relative ratios up to a factor of 5 (or even 10) result in a difference, by~subtraction, 
between~the time intervals of 1.6 (2.3), which correspond to a percent variation by 30\% (50\%) 
only, i.e.,~well within the uncertainties. For~\iso{244}Pu/\iso{238}U, instead, it is more 
difficult to make a quick estimate because of the very different $\tau$. In the next section, 
we evaluate quantitatively, using the WINNET set, the values of the time elapsed between 
production and incorporation into the first solids in the ESS to verify which models can match 
the three constraints~simultaneously.

%For a rough estimate, models with \iso{129}I/\iso{127}I roughly {\bf a few times}  times larger than \iso{247}Cm/\iso{235}U are able to reproduce the ESS values of these isotopes, since \iso{129}I/\iso{127}I in the ESS is roughly twice as high as \iso{247}Cm/\iso{235}U. 
%As for \iso{244}Pu/\iso{238}U, its ESS value is more than 100 times higher than, for example, that of \iso{247}Cm/\iso{235}U (and ln(100)$\sim$4.6) but $\tau_{244}$ is also 3.6 times higher than $\tau_{247}$, therefore, models with \iso{244}Pu/\iso{238}U roughly 30\% (4.6/3.6) higher than \iso{247}Cm/\iso{235}U are expected to also match this constrain. 

%{\bf I think need the figures with the reference isotopes yields relative to solar, also for the steady state, which models to chose, chat with Benjamin about that. I also need more ticks everywhere :) }

\unskip

\unskip

\section{Galactic Evolution and Origin of the SLRs in the~ESS}
\label{sec:galaxy}

When considering the ESS data, we need to process the stellar abundances for their recycling within the ISM material from which the Sun formed. Such recycling implies a certain time delay, which is crucial to consider when analysing radioactive isotopes that decay within a given timescale. C\^ot\'e~et~al.~\cite{cote19PaperI,cote19PaperII} and Yag\"ue L\'opez~et~al.~\cite{yague21PaperIII} provided a methodology and tools to address the evolution in SLR nuclei in the ISM of the Galaxy, and we base our analysis on such~works. 

First, we need to take into account the uncertainties related to galactic chemical evolution (GCE) itself over the whole lifetime of the Galaxy. These result in a factor $K$, by which any ratio predicted by nucleosynthesis calculations involving a stable or long-lived reference isotope needs to be multiplied. This factor takes into account the history of the Galaxy and how it influences the evolution, and~therefore the abundance, at the galactic time of the formation of the Sun, of~the stable or long-lived isotope that is used as reference for the ESS ratio. The~values of $K$ we calculated from the full GCE models~\cite{cote19PaperI} are reported in Table~\ref{tab:intro} for the three isotopic ratios considered here. Three values are provided: the middle value is the best-fit case and the other two reflect the GCE uncertainties, which provide a minimum and maximum value of the SLR to stable or long-lived isotope ratios. 
{Summarizing %MDPI:we remove the bold format, please check.%ML: thanks!
 Table~1 of~\cite{cote19PaperI}, the~GCE parameters that mostly affect the value of $K$ are those related to the first and second infall episodes ($A_1$ and $A_2$) and the star formation efficiency ($f_{star}$). The~observational constraints whose uncertainties affect $K$ the most are the current inflow rate and mass of gas. Due to the feedback between all these quantities, there is not a simple relation with the value of $K$. For~example, the~$K_{\rm max}$ values are found for the highest values of $A_2$, $f_{star}$, and~inflow rate, together with the lowest values of $A_1$ and mass of gas. 
The reasons for this behaviour are explained in detail in~\cite{cote19PaperI}.}

We found that, if the reference isotope is stable, as~in the case of \iso{127}I, the~best-fit value of $K$ is 2.3. When the reference isotope is unstable and long-lived (such as \iso{235,238}U), the~value of $K$ decreases with the half life of the nucleus because the abundance is affected by a shorter time scale within the full history of the Galaxy. For~example, in~the case of \iso{235}U, with~a half life of $\sim 1$ Gyr, i.e.,~roughly ten times shorter than the age of the Galaxy, the~$K$ factor decreases by roughly a factor of two.
In the case of  \iso{129}I/\iso{247}Cm, there are no values of $K$ to be applied (in other words, $K$ is always equal to 1) because these two isotopes are both short-lived and insensitive to the past history of our Galaxy. This is one the several advantages of using such a ratio, as~discussed in detail in Paper~I.

The other potential problem is that injection of SLR nuclei into the ISM by the stellar objects that produce them is not continuous, because stellar ejection events happen in correspondence to very specific discrete events, e.g.,~supernova explosions or neutron--star %Please check intended meaning has been retained
mergers. For~stable nuclei, this effect is not significant because their ESS abundances are primarily defined by the total number of events that enriched the pre-solar nebula, rather than by the exact times at which the events occurred. However, for~SLR nuclei, this effect can completely control their abundances in the ISM since they freely decay between events. One way to account for this is to consider the average of the interval $\delta$ between additions to a given parcel of ISM gas from events of a given type, and~compare it to the mean-life $\tau$ of the SLR nuclei produced by this type of events. Therefore, the~$\tau/\delta$ ratio is the crucial parameter to consider, or~equivalently $\tau/\gamma$, where $\gamma$ is the time interval between the births of the event progenitors\endnote{Since $\gamma \simeq <\delta>$ (see detailed discussion in~\cite{cote19PaperII}) for our purposes here $\gamma$ will be considered equivalent to $\delta$.}. We do not know a priori the value of $\tau/\gamma$ for any SLRs and their sources because it depends on uncertain effects such as diffusive transport in the ISM, supernova energetic in carrying material in the ISM, spatial distribution of the events, and~distance of the events from the pre-solar ISM parcel of gas (see, e.g.,~~\cite{banerjee22} and Wehmeyer~et~al., in~prep). 
Our approach has therefore been to first develop a general framework and then test its implications and derive its predictions for different values \mbox{of $\tau/\gamma$. }

C\^ot\'e~et~al.~\cite{cote19PaperII} found that, if $\tau/\gamma>2$ where $\gamma$ is the time interval between the births of the event progenitors, then we can treat the injection of SLR from such an event as continuous {(hereafter %MDPI: we remove the bold format, please check.%ML: thanks!
 Regime I)}. We just need to add an uncertainty resulting from the statistical spread of the SLR abundance. If, instead, $\tau/\gamma<0.3$, the~most likely scenario is that the ESS abundance of the given SLR came from one last event only, without~any memory of the previous events { (hereafter Regime III). For~values of $\tau/\gamma$ in-between 0.3 and 2, the~SLR abundance carries the memory of a few events (hereafter Regime II)}. Finally, we note that considering SLR ratios such as the \iso{129}I/\iso{247}Cm ratio in the last row of Table~\ref{tab:intro} significantly reduces the uncertainties resulting from the discrete nature of stellar ejections, especially for cases when the half lives are comparable, as~discussed in general in Ref.~\cite{yague21PaperIII}, and~in detail for the $r$-process SLR in Paper~I.

In the following, we use the WINNET abundances to derive more information on the early Solar System and the history of presolar matter from the three $r$-process isotopes considered here in different possible scenarios. 
%To verify the first order effect of the nucleosynthetic production on matching the ESS abundances, here, we remain within the one event scenario for \iso{129}I and \iso{247}Cm and apply different, but complementary, possibilities to the origin of \iso{244}Pu. 
%The aim is to investigate possible self-consistent solutions within the current $r$-process model predictions that could account for all the four isotopic ratios reported in Table~\ref{tab:intro}. 
We remind that the main difference between \iso{129}I and \iso{247}Cm, on~the one hand, and~\iso{244}Pu, on~the other hand, is that the half life of the latter is roughly five times longer than those of the former two. Therefore, the~$\tau/\gamma$ criterion needs to be applied differently, even if all the three isotopes are exclusively produced by $r$-process~events.

\subsection{One (Regime III) or Few (Regime II) Events and Time Elapsed from Last Event}
\label{sec:last}

\begin{figure}[H]

\begin{adjustwidth}{-\extralength}{0cm}
\centering %% If there is a figure in wide page, please release command \centering
\includegraphics[width=9 cm]{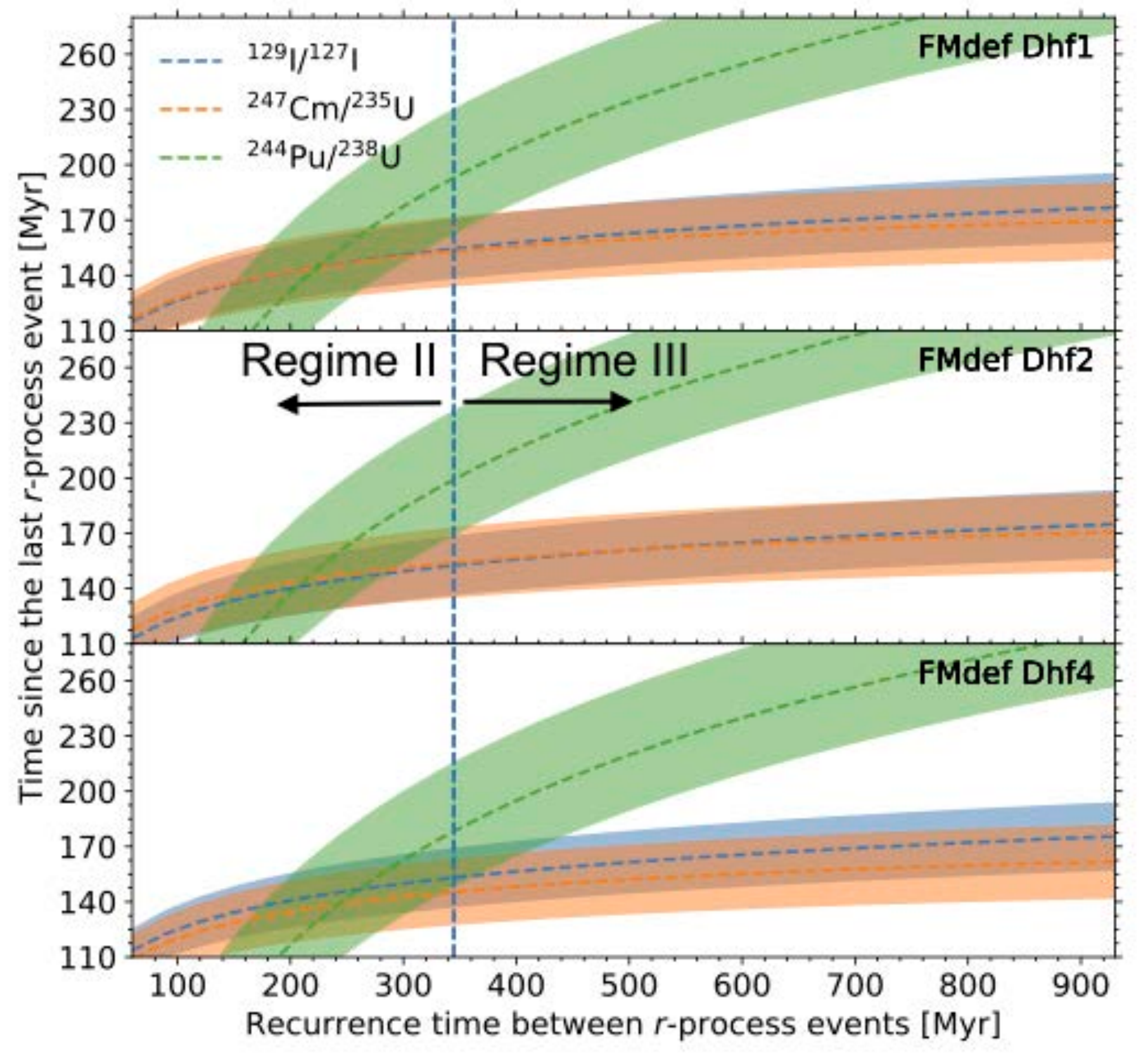}
\includegraphics[width=9 cm]{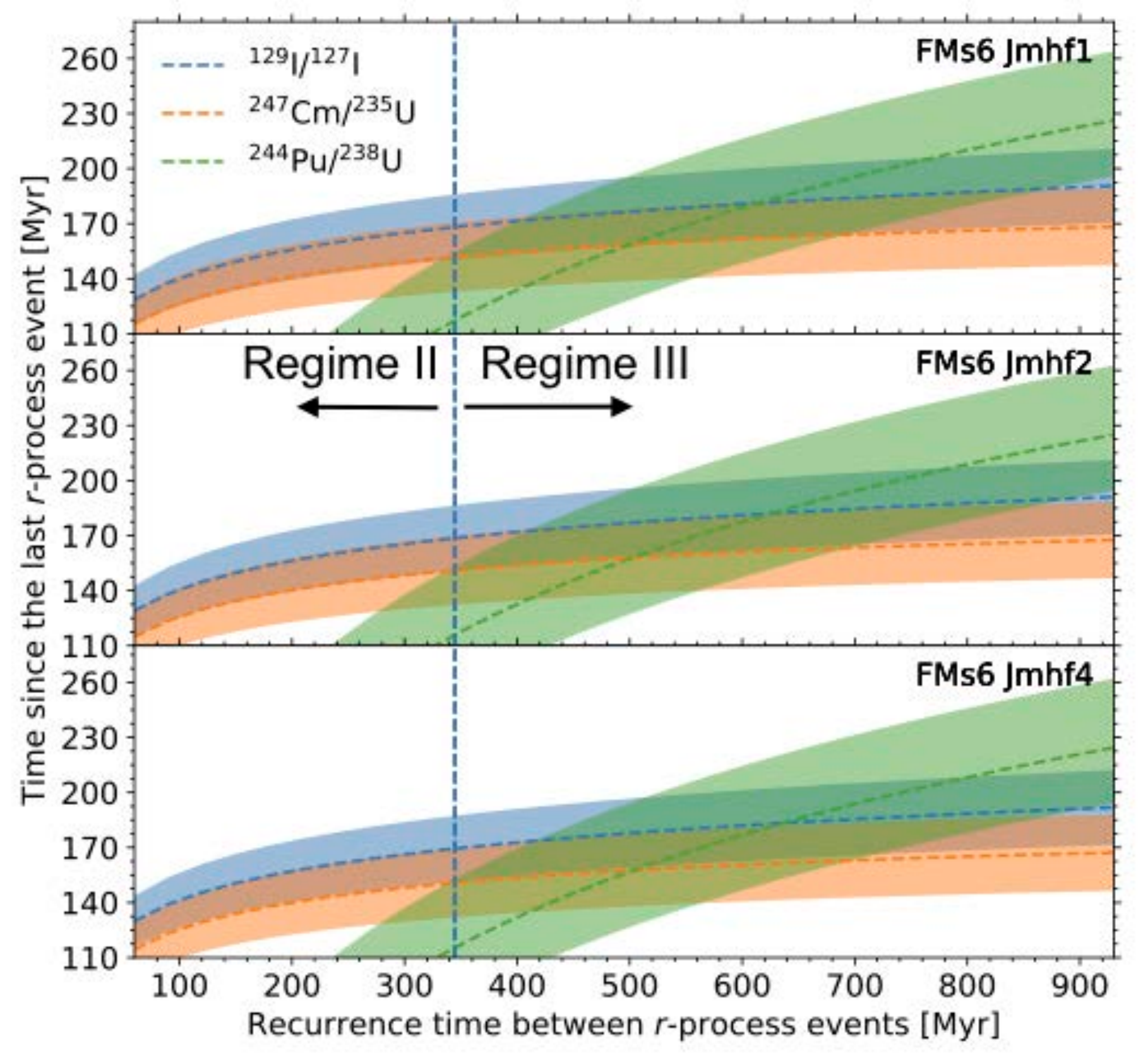}
\end{adjustwidth}
%\begin{figure}[H]
%\includegraphics[width=9.5 cm]{lastevent/FMs6 Fmhf1_best.pdf}
\caption{Examples of solutions obtained using {the %MDPI: we remove the bold, please confirm.%ML thanks
 WINNET set of yields and} $K_{\rm best}$ for the time elapsed from the last $r$-process event to the formation of the first solids in the ESS that are consistent for all the isotopic ratios of Table~\ref{tab:intro}, except~for the right top and middle panels, which correspond to \iso{129}I/\iso{247}Cm ratios just outside the required range.
The colored dashed lines show the time elapsed as function of the free parameter $\delta$ derived from each ratio (labels in the top left corner). 
The dashed blue vertical line represent the $\delta$ value 345 Myr for which $\tau_{244}/\delta=0.3$, {which marks the border between Regimes II and III.}
Uncertainty bands are a composition of the error distributions around the $\tau$ and the ESS ratio for each isotope. They are calculated using a Monte-Carlo sampling of such error distributions (using the 1$\sigma$ values and normal distributions, as~required). The~plotted uncertainty bands are the 2$\sigma$ uncertainty of the Monte-Carlo runs. (Note that these areas are smaller than those shown in Figure~S2 of Paper I because there all the different values of $K$ where included in the bands.)
\label{fig:lastevent}}
\end{figure}

In the case of the two $r$-process SLR \iso{129}I and \iso{247}Cm, as~discussed and presented in detail in Paper I, we can justify statistically the assumption that their abundances in the ESS originated from one last event only, {(Regime III %MDPI: we remove the bold format, please check.%ML OK! thanks!
)}, which occurred roughly 100-200 Myr before the formation of the first solids in the Solar System. The~criterion $\tau/\gamma<0.3$ under which { Regime III} is valid {also for \iso{244}Pu is} that $\gamma$, or~equivalently $\delta$ in the equation, is greater than 345 Myr. Therefore, possible solutions for this scenario {are} those for which $\delta$ is around or larger than this value. To~evaluate the ISM \iso{244}Pu/\iso{238}U ratio under the assumption that \iso{129}I, \iso{247}Cm, and~\iso{244}Pu in the ESS originated from one event, we then use Eq.~S2 of Paper I, as~performed in that paper for \iso{129}I/\iso{127}I and \iso{247}Cm/\iso{235}U, and~the values of $K$ reported in Table~\ref{tab:intro}. Some examples of {the calculation of the time from the last event} are shown in Figure~\ref{fig:lastevent}. There, self-consistent solutions are represented by the overlapping areas of the three different colored bands, each representing one of the three SLR isotopes and their uncertainties. {The trend  with the $\delta$ of the time elapsed calculated using \iso{244}Pu is steeper than those calculated using the other two isotopes. This is due to its much longer $\tau$ value and the fact that the time elapsed is a linear function of $\tau$.}
%The abundances used in this figure are taken from the WINNET set, and we note the the three FMdef Jhf cases produce almost equivalent results to those shown in the left panel (i.e., the three FMdef Dhf cases). 

{Figure~\ref{fig:lastevent} also shows some examples of possible solutions for Regime II, which corresponds to} $0.3<\tau/\delta<2$, i.e.,~$\delta$ = 68\endnote{This lower limit is defined such that $\tau/\delta<0.3$ for \iso{129}I and \iso{247}Cm, but~it is close to the 57.5 Myr value defined by $\tau/\delta>2$ for \iso{244}Pu.}- 345 Myr. {In this case,} \iso{244}Pu originated from {a few discrete} events and the lower the value of $\delta$, the~larger the number of events. The~last event would have contributed only a fraction, {$1 - e^{(-\delta/\tau)}$ (assuming a constant production factor),} of the ESS abundance of \iso{244}Pu. {Therefore, at~the lower limit of Regime II, $\delta=68$ Myr, the~last event contributed 45\% of the ESS abundance of \iso{244}Pu.}

%{\bf in the figure here the Fmhf is Jmhf in the table above ...(?) is that a typo maybe?
%Also Andres, would it be much work to have these figures down to delta=100? just because it would help wit the discussion at Section~3.2}

Overall, the~WINNET set comprises 63 sets of models, and~the GCE model provides three values of $K$ for a total of 189 possibilities. For~the first three ratios of Table~\ref{tab:intro}, we found {that 
%MDPI: we remove the bold format, please check..%ML thank, OK
 92\% of the models can provide overlapping solutions: 62 of those with $K_{\rm min}$, 60 of those with $K_{\rm best}$, and~52 of those with $K_{\rm max}$.}
%58 overlapping solutions using $K_{\rm min}$, 41 using $K_{\rm best}$ and 8 using $K_{\rm max}$. So a total of 107 cases provide a solution (i.e., 57\% out of all 189 models), 
Therefore, solutions are common, {partly thanks to the degree of freedom provided by the relatively free parameter $\delta$}. Times of the last event {are} in the range 100-200 Myr as derived in Paper I, {this is expected given that the analysis presented here is} just an extension of that presented there, to check if the \iso{244}Pu/\iso{238}U ratio could also be explained. {A new result is that these elapsed times are lower in Regime II than in Regime III, due to the steeper trend with $\delta$ of those calculated using \iso{244}Pu.}

If we include the requirement that the \iso{129}I/\iso{247}Cm ratio should be between 254 and 622\endnote{Note that the evaluation of the ESS ratio of \iso{129}I/\iso{247}Cm depends on the time from the last event itself, given that its $\tau_{\rm ratio}$ is variable due to the uncertainties in $\tau_{\rm 129}$ and $\tau_{\rm 247}$, as~discussed in Section~\ref{sec:intro}. The~ESS values reported in Table~\ref{tab:intro} and used here were calculated assuming a time from last event in the range 100-200 Myr and composing all the uncertainties, as~discussed in detail in the supplementary material of Paper I. A~more precise analysis would instead use the range of times from the last event for each model solution to derive the range of corresponding ESS \iso{129}I/\iso{247}Cm ratios, and~find if the model matches such a specific range. However, given that, as~noted above, most solutions provide times in the 100-200 Myr range, this more accurate treatment would not change our results.}, the~number of solutions becomes much more restricted.
%Interestingly, while the times elapsed are in any case in the range 140-170 Myr, all the solutions obtained using FMdef have $\delta$ values ranging from the limit of what we have assumed in order for \iso{244}Pu to originate from one last event (345 Myr, blue vertical line in the figure), up to roughly 400 Myr at most. For the solutions from the FMs6 set, instead, $\delta$ reaches up to $\simeq$900 Myr, which would imply that only roughly ten $r$-process events in the history of the Galaxy produced the $r$-process inventory of the Solar System.  The $K_{\rm best}$ is used in all the examples shown in the figure, but the plots produced using $K_{\rm min}$ are almost equivalent, except that the times are slightly shorter (as the lower $K$ results in lower ratios and therefore shorter times). 
{In fact,} the \iso{129}I/\iso{247}Cm ratio is a much more stringent constraint because
the two isotopes are very far apart in mass and therefore located in very different regions of the nuclide chart. Such relative abundances are more sensitive to the general features of the process and its astrophysical site (such as the amount of free neutrons) as well as the uncertainties in the nuclear model, than~the ratios of isotopes that are closer to each other in mass (see Figure~\ref{fig:ratios}).
As shown in Figure~S2 of Paper I, the~simulations that produce the best matches to the observations are those dominated by moderately neutron-rich ejecta (in the specific case of the WINNET models, these correspond to the nine FMdef and the three FMs6 Jmhf models).

Out of these models, we find that six out of the nine FMdef models (the three Dhf plus the \mbox{three Jhf cases}) and one of the FMs6 model (Jmhf4) can also account for \iso{244}Pu/\iso{238}U {when using either of the three values of $K$, for~21 solutions in total. The~main difference between using  $K_{\rm min}$ and $K_{\rm best}$ versus using $K_{\rm max}$ is that the former two values provide solutions for $\delta$ values typical of Regime III, while the latter corresponds to solutions within Regime II.}  
%Also one of the FMs6 model (Jmhf4) provides a solution for any of the three values of $K$, so that the total number of solutions is 15. 
(The FMs6 Jmhf1 and Jmhf2 cases produce \iso{129}I/\iso{247}Cm ratios of 236 and 242, respectively, just outside the required range). In~summary, {more than half (21)} of all the 36 possible models that match the \iso{129}I/\iso{247}Cm (12 models $\times$ 3 values of $K$ = 36) provide a global solution for all the four isotopic~ratios. 

Finally, we note that, if the ESS \iso{244}Pu/\iso{238}U ratio was lower than the value used here, the~green shaded area in Figure~\ref{fig:lastevent} would shift upwards, for~example, by~80 Myr if the ESS ratio was twice as low, due to a longer decay time needed to match the lower ESS value. This would remove {most of the Regime III} solutions and shift {the Regime II} solutions to lower values of $\delta$.

\subsection{Steady-State Equilibrium (Regime I) and Isolation Time}
\label{sec:steady}

{In %MDPI:we remove the bold format, please check %ML ok thanks!
 Regime I,} $\tau/\delta$ is $> 2$, {corresponding} to $\delta < 57.5$ Myr, and~\iso{244}Pu evolves in steady-state equilibrium. In~this case, \iso{129}I and \iso{247}Cm would be in {Regime II; the~time elapsed from the last event would decrease with $\delta$ (as discussed in Section~\ref{sec:last}), and~reach roughly 80-130 Myr}.  %Therefore, the~time elapsed from the last event would be higher than those reported in Section~\ref{sec:last} and Figure~\ref{fig:lastevent} because this last event would have contributed only a fraction of their ESS abundance (similarly as discussed analysed in Section~\ref{sec:middle} for the case of \iso{244}Pu). 
The steady-state regime for \iso{129}I and \iso{247}Cm would require, instead, roughly $\delta < 11$ Myr {(for this value of $\delta$, at~the limit of Regime II, the~last event contributed roughly 40\% of their ESS abundances)}. This can be excluded with reasonable confidence because it is the typical value obtained for core-collapse supernovae, which are much more frequent than the currently accepted $r$-process~sources. 

If \iso{244}Pu was in steady-state equilibrium, then we can use Equation~(11) of~\cite{lugaro18rev} (where $K=k+1$) and consider the productions ratios from the $r$-process models as a continuous wave of enrichment. In~this case, the~time interval needed to decay the ISM ratio to its corresponding ESS ratio is  
an isolation time rather than a time from the last event. This time interval can then be compared to the isolation time obtained from the $s$-process isotopes, \iso{107}Pd, \iso{135}Cs, and~\iso{182}Hf, under~the assumption of same regime, i.e.,~$\delta < 5-6$ Myr for the $s$-process events in the Galaxy, which correspond to asymptotic giant branch (AGB) stars of initial mass $\simeq2-4$ \msun\ ~\cite{trueman22}. For~the three values of $K$ to be used when studying SLR/stable isotope ratios (i.e., 1.6, 2.3, 5.7, as~reported in Table~\ref{tab:intro} for \iso{129}I/\iso{127}I), the~isolation times reported by~\cite{trueman22} for the $s$-process SLR isotopes are 9-12, 10-16, and~18-26 Myr, respectively. We also need to consider the statistical uncertainty due to stochasticity and discussed in~\cite{cote19PaperII}. We can use here the uncertainties reported in Table~3 of~\cite{cote19PaperII}, for~the specific case $\tau/\gamma=3.16$ and $\gamma$=31.6 Myr, which are close to the maximum uncertainty that would correspond to the case of \iso{24}Pu in this regime. The~error is almost symmetric and corresponds to variations in the ISM ratio of +1.16 and $-$0.84. These translates into error bars to be applied to each isolation time of $+$17 and $-$19~Myr.

In the case of $K_{\rm max}$, no solutions are present because all the isolation times derived from \iso{244}Pu are in the range 61-213 Myr, much higher than the range derived for the $s$-process SLR nuclei of 18-26 Myr. This is controlled by the large value of $K_{\rm max}$ combined with the long half life of \iso{244}Pu. In~the case of $K_{\rm min}$ and $K_{\rm best}$, instead, there are 18 and 14 solutions possible, respectively, which have an overlap with the ranges of isolation time derived from the $s$-process SLR nuclei. These solutions are all obtained from the models that produce \iso{244}Pu/\iso{238}U abundance ratios in the range 0.19--0.34. As~shown in  Figure~\ref{fig:ratios}, these correspond mostly to the WINNET models run with the Jmhf nuclear inputs (out of the total 32 solutions, 23, , i.e.,~72\%, are Jmhf solutions) and the six NS--NS merger PRISM models with SLY4, TF\_Mkt, and~UNEDF0. For~the other nuclear input choices, instead, only specific astrophysical sites results in \iso{244}Pu/\iso{238}U abundance ratios in the required~range. 

Out of the 12 models that match the three ratios that involve \iso{129}I and \iso{247}Cm, seven of them also provide solutions for the isolation time from \iso{244}Pu/\iso{238}U compatible with the $s$-process SLR isotopes. However, as~mentioned above for the value of $\delta$ considered here, these two SLR isotopes may have more than one event contributing to their ESS abundances; therefore, such constraints become less strong (see also~\cite{banerjee22}). 

We should also consider the case where \iso{244}Pu is in steady-state, but~the $s$-process SLRs came from one last event, which requires roughly $\delta >30$ Myr for $s$-process event in the Galaxy. In~this case, the~last $s$-process event was identified to have occurred at 25 Myr before the formation of the first solids~\cite{trueman22}, therefore, the~isolation time from \iso{244}Pu/\iso{238}U is simply constrained to be smaller than this value. Also in this case solutions do not exist with $K_{\rm max}$, while there are 13 and 7 more solutions for $K_{\rm min}$ and $K_{\rm best}$, respectively. Most of these solutions overlap as they correspond to the same $r$-process models but the different value of $K$, therefore, they correspond to the same range of \iso{244}Pu/\iso{238}U abundance ratios and nuclear models as reported above. The~isolation time from \iso{244}Pu in this case can vary more more freely and there are a few models that given the uncertainties provide values down to zero, which is not a useful constraint. Finally, we note that if the $s$-process SLRs originated from a few events (i.e., $5-6<\delta <30$ Myr), then the time from the last event would increase and a few more models could produce an isolation time lower than this value. A~more detailed statistical analysis would be needed in this~case.

Finally, we note that if the ESS value of the \iso{244}Pu/\iso{238}U ratio was twice as high as the value considered here, we would need to add $+80$ Myr to every isolation time, which would make it impossible to find a solution consistent with the origin of the $s$-process \mbox{SLR isotopes.}

\section{Summary and~Conclusions}
\label{sec:conclusions}

We presented and analysed the relative production of the short-lived and long-lived $r$-process isotopes \iso{129}I, \iso{235}U, \iso{238}U, \iso{244}Pu, and~\iso{247}Cm and the stable \iso{127}I in a large set of 119 $r$-process models from two different sets calculated with the WINNET and PRISM frameworks. We then investigated if it is possible to find solutions for the origin of the ESS abundance of \iso{244}Pu that provide production at the source and time intervals (either from the last event or from the time of the isolation, depending on the $\tau/\delta$ regime) compatible to those of the other $r$-process and $s$-process SLR isotopes. {A summary of the different possibilities, solutions, and~derived time intervals are shown in Table~\ref{tab:summary}. In~brief:}
\begin{table}[H]
\caption{Summary of the different regimes combinations for the different SLRs, their corresponding $\delta$ values in Myr ($\delta_r$ and $\delta_s$, for~the $r$- and $s$-process events, respectively), $r$-process model solutions, and~elapsed time ($t_{{\rm e},r}$ and $t_{{\rm e,}s}$, for~the last $r$- and $s$-process event, respectively) or isolation time ($t_{\rm i}$) in Myr. Notes: $^{a}$For all the four ratios in Table~\ref{tab:intro}: 6 = FMdef(3xDhf,3xJhf) + FMs6Jmhf4, all valid for each of the three values of $K$. We did not check the PRISM %Please check intended meaning has been retained
	models for these regimes. $^{b}$ Of which 23 have Jmhf nuclear input. $^{c}$The two NS--NS merger models with the three nuclear inputs: SLY4, TF\_Mkt, and~UNEDF0.  \label{tab:summary}}
%\newcolumntype{C}{>{\centering\arraybackslash}X}

\begin{adjustwidth}{-\extralength}{0cm}
%\centering %% If there is a figure in wide page, please release command \centering
\setlength{\cellWidtha}{\fulllength/4-2\tabcolsep+0.9in}
\setlength{\cellWidthb}{\fulllength/4-2\tabcolsep-0.3in}
\setlength{\cellWidthc}{\fulllength/4-2\tabcolsep-0.3in}
\setlength{\cellWidthd}{\fulllength/4-2\tabcolsep-0.3in}
\scalebox{1}[1]{\begin{tabularx}{\fulllength}{>{\centering\arraybackslash}m{\cellWidtha}>{\centering\arraybackslash}m{\cellWidthb}>{\centering\arraybackslash}m{\cellWidthc}>{\centering\arraybackslash}m{\cellWidthd}}
\toprule

\textbf{Regime} & \textbf{$\delta$ (Myr)} & \textbf{Solutions} & \textbf{Times (Myr)} \\ 
\midrule
III for \iso{129}I, \iso{247}Cm, and~\iso{244}Pu & $\delta_r$ $>$ 345 & \multirow{2}{*}{7 WINNET$^{a}$} & \multirow{2}{*}{$t_{{\rm e},r}$ $\simeq$ 100--200}\\
III for \iso{129}I and \iso{247}Cm and II for \iso{244}Pu & 68 $<$ $\delta_r$ $<$ 345 & & \\
\midrule
II for \iso{129}I and \iso{247}Cm and I for \iso{244}Pu, \iso{107}Pd, and~\iso{182}Hf & 11 $<$ $\delta_r$ $<$ 68, $\delta_s$ $<$ 5 & 32 WINNET$^{b}$, 6 PRISM$^{c}$, 0 for $K_{\rm max}$ & \multirow{2}{*}{$t_{{\rm e},r}$ $\simeq$ 80--130, $t_{\rm i}$ $\simeq$ 9--16} \\ 
%with I for \iso{107}Pd and \iso{182}Hf & $\delta_s$ $<$ 5 & & \\
OR III for \iso{107}Pd and \iso{182}Hf & 11 $<$ $\delta_r$ $<$ 68, $\delta_s$ $>$ 30 & 20 more than above & $t_{{\rm e},s}$$\simeq$ 25, $t_{\rm i}$ > 0 \\
\bottomrule
\end{tabularx}}
\end{adjustwidth}
\end{table}
\begin{enumerate}
    \item In {Section~\ref{sec:last} (top section of Table~\ref{tab:summary}), we considered Regimes II and III for} \iso{244}Pu, {corresponding to $\delta > 68$ Myr and Regime III for} \iso{129}I and \iso{247}Cm. {More than half} of the WINNET models that were already shown to reproduce the three ratios that involve \iso{129}I and \iso{247}Cm in Paper I, also provide a self-consistent solution for \iso{244}Pu. These models all correspond to the NS--NS merger disk cases dominated by moderately neutron-rich ejecta.
    %Overall, models that use $K_{\rm min}$ or $K_{\rm best}$ to represent the effect of the full history of the Galaxy on evolution of the ratios considered provide 92\% of the solutions for a self-consistent time for the last event between 100 and 200 Myr before the birth of the Sun. Out of the 15 models that also match the \iso{129}I/\iso{247}Cm ratio, only one of them has $K_{\rm max}$.
%    \item The cases of \iso{244}Pu in the ESS originating from a few events instead of just one and/or an ESS value of \iso{244}Pu twice higher than the value used here can also be accommodated by the models discussed above in Point 1, with the main difference being that the values of the time interval between different injections, $\delta$, would be somewhat lower.
    \item In {Section~\ref{sec:steady} (bottom section of Table~\ref{tab:summary}), we considered Regime I for \iso{244}Pu, i.e.,~$\delta < 68$ Myr}, where this SLR reaches a steady-state value in the ISM. It is also possible to find a significant number of $r$-process models (mostly corresponding to the Jmhf nuclear input) that provide solutions for the ESS \iso{244}Pu abundance compatible to those of the SLR isotopes produced also by the $s$ process: \iso{107}Pd and \iso{182}Hf (and the current ESS upper limit of \iso{135}Cs). However, no solutions exist {in Regime I for \iso{244}Pu} if the ESS value of \iso{244}Pu was twice as high as the value used here or if the Milky Way model was represented by $K_{\rm max}$. 
\end{enumerate}

We cannot determine if the solution to the origin of \iso{244}Pu in the ESS is 1. or 2. above, {and which implications on the timescales are valid,} since we still do not know how far material from $r$-process sources can travel, and, therefore, how many parcels of the ISM are affected by each of these events and the value of $\delta$. {Although we note that recent hydrodynamical models aimed at calculating how far material travels after being ejected by a hypernova predict relatively short distances~\cite{amend22}, which would support large $\delta$ values for the r-process events.}
{In any case, we have established that within Point 1., WINNET solutions within the NS--NS disk ejecta favour the Dhf and Jhf nuclear models. Within~Point 2., all} solutions exclude the case of a Milky Way Galaxy with $K_{\rm max}$, therefore restricting {the isolation time} to 9-16 Myr {(if \iso{107}Pd and \iso{182}Hf are also in Regime I),} still supporting the hypothesis that the Sun was born in a massive, long-living molecular cloud. We can also conclude that a much lower value of the \iso{244}Pu/\iso{238}U ratio in the ESS than that reported in Table~\ref{tab:intro} would be impossible to reconcile {within Regime I, and~would therefore support Regimes II and III}. New, future experiments and analysis are needed to confirm the ESS \iso{244}Pu/\iso{238}U ratio.
%This results is still depending on the assumption that the $s$-process SLR are in steady-state equilibrium, which is also affected by the value of the $\delta$ for the $s$-process sources. 

%The intermediate neutron-capture process (the $i$ process, e.g.,~~\cite{hampel16}) may also lead to their production, but has not been investigated yet. 
\vspace{6pt}

\supplementary{The following supporting information can be downloaded at: \linksupplementary{s1}, 
Table S1: WINNET-abundances.txt; Table S1: PRISM-ratios.txt %MDPI: please add the title of suppl.

 }

\authorcontributions{All %MDPI: please check the author: B.C.; M.P.; N.V.; B.W.; M.P. are not 
mentioned in this part, please add
 the authors have contributed to the conceptualization, methodology, software, validation, formal analysis, 
and~investigation. The~original draft was prepared by M.L. %MDPI: The abbreviation of M\'aria Pet\H{o} 
and Marco Pignatari are the same, please indicate them
 with Figure~\ref{fig:ratios129247} contributed by B.S., and~Figure~\ref{fig:ratios} by A.Y.L.. A.Y.L., B.S., 
B.C., M. Pet\H{o}, N.V., B.W., and M. Pignatari  
contributed to the methodology, as well as the review and editing of the paper. M.L. contributed to the supervision, project 
administration, and~funding acquisition for the project. All authors have read and agreed to the published 
version of the~manuscript.}

\funding{This research was funded by ERC via CoG-2016 RADIOSTAR (Grant Agreement 724560). The~work of AYL 
was supported by the US Department of Energy through the Los Alamos National Laboratory. Los Alamos 
National Laboratory is operated by Triad National Security, LLC, for~the National Nuclear Security 
Administration of U.S.\ Department of Energy (Contract No.\ 89233218CNA000001). BC acknowledges support 
of the National Science Foundation (USA) under grant No. PHY-1430152 (JINA Center for the Evolution of 
the Elements).} %ML added one acknoledgement for BC

\institutionalreview{Not applicable %MDPI: In this section, you should add the Institutional Review Board 
Statement and approval number, if~relevant to your study. You might choose to exclude this statement if 
the study did not require ethical approval. Please note that the Editorial Office might ask you for 
further information. Please add “The study was conducted in accordance with the Declaration of Helsinki, 
and~approved by the Institutional Review Board (or Ethics Committee) of NAME OF INSTITUTE (protocol code 
XXX and date of approval).” for studies involving humans. OR “The animal study protocol was approved by 
the Institutional Review Board (or Ethics Committee) of NAME OF INSTITUTE (protocol code XXX and date of 
approval).” for studies involving animals. OR “Ethical review and approval were waived for this study due 
to REASON (please provide a detailed justification).” OR “Not applicable” for studies not involving 
humans or animals..
j}

\informedconsent{Not applicable %MDPI: Any research article describing a study involving humans should contain 
this statement. Please add ``Informed consent was obtained from all subjects involved in the study.'' OR 
``Patient consent was waived due to REASON (please provide a detailed justification).'' OR ``Not 
applicable'' for studies not involving humans. You might also choose to exclude this statement if the 
study did not involve humans.Written informed consent for publication must be obtained from participating 
patients who can be identified (including by the patients themselves). Please state ``Written informed 
consent has been obtained from the patient(s) to publish this paper'' if applicable..
}

\dataavailability{Not applicable %MDPI: In this section, please provide details regarding where data supporting 
reported results can be found, including links to publicly archived datasets analyzed or generated during 
the study. Please refer to suggested Data Availability Statements in section ``MDPI Research Data 
Policies'' at \url{https://www.mdpi.com/ethics}. If~the study did not report any data, you might add 
``Not applicable'' here..
} 

\acknowledgments{We thank Marius Eichler, Almudena Arcones, and~Thomas Raucher for providing us with the WINNET models and their results of $r$-process nucleosynthesis. We thank Matthew Mumpower, Trevor Spouse, and~Rebecca Surman for their contributions to the PRISM models. We also thank Jamie Gilmour for discussion on the ESS data.
MP acknowledges support of NuGrid from NSF grant PHY-1430152 (JINA Center for the Evolution of the Elements) and STFC (through the University of Hull's Consolidated Grant ST/R000840/1), and~access to {\sc viper}, the~University of Hull High Performance Computing Facility. MP acknowledges the support from the ``Lendület-2014'' Programme of the Hungarian Academy of Sciences (Hungary). We thank the ChETEC COST Action (CA16117), supported by COST (European Cooperation in Science and Technology), and~the ChETEC-INFRA project funded from the European Union’s Horizon 2020 research and innovation programme (grant agreement No 101008324), and~the IReNA network supported by NSF AccelNet.}

\conflictsofinterest{The authors declare no conflict of interest. The~funders had no role in the design of the study; in the collection, analyses, or~interpretation of data; in the writing of the manuscript, or~in the decision to publish the~results.} 

\newpage
\abbreviations{Abbreviations}{
The following abbreviations are used in this manuscript:\\

\noindent 
\begin{tabular}{@{}ll}
ESS & early Solar System\\
GCE & galactic chemical evolution\\
ISM &  interstellar medium \\
NSM & neutron star merger \\
$r$ process  &  $rapid$ neutron-capture process \\ %Please check intended meaning has been retained

$s$ process &  $slow$ neutron-capture process \\ %Please check intended meaning has been retained

SLR & short-lived radioactive 
\end{tabular}}

\begin{adjustwidth}{-\extralength}{0cm}
\printendnotes[custom]

\reftitle{References}
%\externalbibliography{yes}
%\bibliography{apj-jour.bib,marialib1.bib,marialib2.bib,marialib3.bib}

\end{adjustwidth}

\end{document}